# Cosmogenic nuclide enhancement via deposition from long-period comets as a test of the Younger Dryas impact hypothesis.

**Andrew C. Overholt[a,b], Adrian L. Melott[a*]**

[1] *Department of Physics and Astronomy, University of Kansas, Lawrence, Kansas 66045*
[2] *Department of Science and Mathematics, MidAmerica Nazarene University, Olathe, Kansas 66062*

**ABSTRACT**

We explore the idea that detectable excursions in $^{26}$Al may arise from direct deposition by any bolide, and excursions in $^{14}$C and $^{10}$Be abundances in the atmosphere may result from long-period comet impacts. This is very different from the usual processes of production by cosmic rays within Earth's atmosphere. Long-period comets experience greatly increased cosmic ray flux beyond the protection of the sun's magnetic field. We report the computed amount of $^{14}$C, $^{10}$Be, and $^{26}$Al present on long-period comets as a function of comet mass. We find that the amount of nuclide mass on large long-period comets entering the Earth's atmosphere may be sufficient for creating anomalies in the records of $^{14}$C and $^{10}$Be from past impacts. In particular, the estimated mass of the proposed Younger Dryas comet is consistent with its having deposited sufficient isotopes to account for recorded $^{14}$C and $^{10}$Be increases at that time. The $^{26}$Al/$^{10}$Be ratio is much larger in extraterrestrial objects than in the atmosphere, and so, we note that measuring this ratio in ice cores is a suitable definitive test for the Younger Dryas impact hypothesis, even if the hypothetical bolide is not a long-period comet and/or did not contribute to the $^{14}$C and $^{10}$Be increases.

* Corresponding author Tel. 785-864-3037
Email addresses: acoverholt@mnu.edu, melott@ku.edu

## 1. INTRODUCTION

Impacts from extraterrestrial objects provide a significant threat over long timescales of damage and mass extinction for terrestrial biological systems. As well as directly damaging biological systems, extraterrestrial impacts are known to create long-lived atmospheric effects, such as ash and dust clouds. Little is known of the frequency of such events, and what is known depends greatly on cratering and other geologic data (Raup & Sepkoski 1984, Bland et al. 1998, Jetsu 2011) which do not measure airbursts. The extinction event at the end of the Cretaceous period (KPg boundary) coincides with a large bolide impact proposed to be the primary cause of the extinction event (Alvarez et al. 1980, Schulte et al. 2010).

Impact events are produced by a variety of extraterrestrial objects, including iron meteorites, rocky asteroids, and icy comets. Meteorites are well studied, because they typically leave craters and residual fragments of the initial object. In other cases, the impactor can break up in an airburst when rapid thermal expansion of the object produces fracturing and an apparent explosion of the nucleus before reaching ground level. These events are more difficult to study and leave less evidence than terrestrial impacts, making their frequency largely unknown. Major airbursts release a large amount of energy that may threaten biological systems, as an example, during the Tunguska event, which occurred above Siberia in 1908 (Farinella et al. 2001). The mass of the Tunguska cosmic body is estimated to have been $5 \times 10^7$ kg (Wasson 2003); it produced the largest airburst in recorded history and felled an estimated 80 million trees. An event much larger than Tunguska has been suggested as the possible cause of the Younger Dryas cooling event of the late Pleistocene. Reported evidence for this impact includes peaks in microspherules, high-temperature melt-glass (>2100°C), iridium, osmium[1], polycyclic aromatic hydrocarbons, and nanodiamonds (Firestone et al. 2007, Kurbatov et al. 2010, Israde-Alcantaraz et al. 2012, Bunch et al. 2012, Mahaney et a. 2013), all of which are also present in the KPg boundary layer, and platinum, which is not (Petaev et al. 2013). The impactor for this event was conjectured to be a 1-4+ km wide comet, making this event many orders of magnitude larger than Tunguska (Bunch et al. 2012). However, this hypothesis and the origin of the reported evidence are still controversial (Kerr 2008, Surovell et al. 2009, Boslough et al. 2012, but see Lecompte et al. 2012). In this paper, we do not address this controversy, but rather whether or not the nuclide evidence is consistent with causation by a large impact.

Long-period comets present a unique threat due to their relative unpredictability (Napier and Asher 2009). These comets often begin in the Oort cloud, a spherical arrangement of extraterrestrial ice and dust ~50,000 AU from our sun; from there, they may be perturbed from orbit and fall towards the center of our solar system. This cloud falls within the interstellar medium (ISM), an environment vastly different from the inner solar system. With very little direct measurement data available for these comets, their cosmogenic properties remain unknown. It was recently argued that the K-Pg bolide was most likely a large period comet (Moore and Sharma 2013).

Comets and other bolides entering Earth's atmosphere will deposit detectable residue, as occurred when the KPg boundary impactor deposited a worldwide layer of iridium that was used as a primary argument for its existence (Alvarez et al. 1980, Shulte et al. 2010). Melott et al. (2010) showed that cometary airbursts may deposit nitrates and ammonium potentially detectable in ice core data, and these are consistent with data for the Tunguska event. If excess cosmogenic isotopes exist in comets due to increased cosmic ray exposure prior to their impact, then comets may also deposit these in the atmosphere, causing a deviation from background abundances in the record. However, because other processes, such as geomagnetic changes or sudden fluctuations in ISM cosmic-ray flux (Erlykin and Wolfendale 2010) can also produce isotopic excursions, increases in such nuclides alone cannot be counted as evidence of an impactor. On the other hand, extraterrestrial matter produces cosmogenic nuclides in different proportions than that of cosmic ray showers in Earth's atmosphere, and therefore, variance in the ratios of cosmogenic nuclides can potentially provide evidence for past impact events. In particular, we show one such ratio to be a good test for the Younger Dryas impact hypothesis.

---

[1] Sharma M, Chen C, Jackson BP, Abouchami W. (2009) High resolution Osmium isotopes in deep-sea ferromanganese crusts reveal a large meteorite impact in the Central Pacific at 12 ± 4 ka. American Geophysical Union, Fall Meeting 2009, abstract #PP33B-06.

## 2. COMET COMPOSITION

In order to compute the expected cosmogenic isotope production on a comet, we make assumptions about its composition, which varies greatly depending on the origin and history of the comet. The comet nucleus (often called a "dirty snowball" or "icy dust ball" depending upon composition) contains rock, frozen gases, and a large amount of water ice combined with ices of other volatiles. The densities of most short-period comet nuclei vary from 0.3 to 0.6 g/cm$^3$ (Britt et al. 2006), whereas the density of long-period comets are probably lower, but largely unknown due to limited measurement opportunities.

Cosmogenic nuclide production occurs when high energy particles collide with atomic nuclei, and these collisions produce secondary particles of varying types in cascades known as a particle showers. Important secondary particles include protons, neutrons, and pions. A large variety of nuclides can be produced in the case of large target nuclei (Michel et al. 1997), making the presence of abundant target elements within long-period comets important. Comets are composed of roughly equal parts ice and dust. Hydrogen, oxygen, nitrogen and carbon are the most abundant elements in cometary ice, while oxygen, carbon and silicon are the most abundant elements in cometary dust. We focus on the most commonly measured cosmogenic nuclides $^{10}$Be, $^{14}$C, and $^{26}$Al, which are produced in collisions with carbon, silicon, oxygen and nitrogen. $^{10}$Be and $^{14}$C are the most abundant cosmogenic isotopes in the Earth's atmosphere. $^{14}$C has been measured in tree rings and ice cores, while $^{10}$Be has been measured in ice cores, developing a record of atmospheric abundance for these two isotopes. Production of $^{10}$Be is caused by the direct spallation of oxygen and nitrogen and $^{14}$C is also produced by the absorption of cosmic ray neutrons by nitrogen. Extensive measurements have been made on the composition of Halley's comet, in particular for its nitrogen content; the N/O ratio within Halley was found to be 0.047 within the dust and 0.023 overall (Wyckoff et al. 1991), far below solar levels. This nitrogen is found as ammonia and in organic molecules. $^{26}$Al is produced at high efficiency in extraterrestrial matter, because of the direct exposure of the silicon in comets or asteroids to cosmic rays. On the other hand, terrestrial production of $^{26}$Al is low because very few cosmic rays reach the ground, with the result that most terrestrial production is confined to the atmosphere, where silicon abundance is low. $^{36}$Cl is another well studied cosmogenic nuclide derived from argon. However, with very little precursor argon present in cometary matter, its production should be negligible. Despite the contribution being small, excess $^{36}$Cl has been measured in fossil rat urine at the onset of the Younger Dryas period (Plummer et al. 1997). This suggests the $^{36}$Cl deposition may be measurable if the Younger Dryas impact hypothesis proves correct.

On their first entrance into the inner solar system, comets contain a large fraction of volatiles. These volatiles (such as ammonia) remain frozen in the comet nucleus while the comet is sufficiently far from the Sun, as in the Oort cloud, but, while on approach, the surface temperature of the comet nucleus is raised through radiant heating to the point of sublimation of these frozen volatiles. As a large portion of cometary nitrogen is found in frozen and gaseous ammonia, long-period comets and comets on first approach contain fractionally higher amounts of nitrogen. For this reason, we will assume the Halley nitrogen ratio as a lower bound; this nitrogen abundance lies well below solar values (Krankowsky 1991, Wyckoff et al. 1991). In addition to nitrogen depletion through outgassing, depletion is also thought to occur during the creation of comets (Iro et al. 2003). Objects originating in low-temperature environments, such as long-period comets from the Oort cloud, may contain more nitrogen than other comets, consistent with solar abundances (Owen 2003). Measurements of distant comets within the heliosphere show higher nitrogen abundances (Korsun et al. 2008). Therefore, we use solar nitrogen abundances as an upper bound.

## 3. COSMIC RAY ENVIRONMENT

Current terrestrial cosmic ray flux is well measured and known, but is not so well known outside the protection of the Earth's magnetosphere, where long-period comets originate. Measurements of the cosmic ray flux in interplanetary space have been taken by the Voyager space probes since launch (Stone et al. 1977), but the probes have yet to completely exit the protection of the Sun's magnetic field. Even so, Voyager is currently entering the interstellar medium and it provides important measurements that serve as benchmarks in choosing a cosmic ray spectrum. The cosmic ray flux of interplanetary space is greater than terrestrial levels, and the flux is also expected to be greater beyond the heliosphere. Many models have been proposed for describing this spectrum (e.g. Spitzer and Tomasko 1968, Hayakawa et al. 1961, Nath and Biermann 1994, Mori 1997, Valle et al. 2002, Kneller et al. 2003, Herbst and Cuppen 2006, Indriolo et al. 2009). Most of these models are now eliminated as the Voyager probe has already detected cosmic ray fluxes in excess of their predictions. The plausible models are those which contain cosmic ray fluxes equal to or exceeding the latest Voyager results (Stone et al. 2013). The recent Voyager results display cosmic ray fluxes in the absence of solar cosmic rays at its location near the heliopause. However, it is uncertain whether GCRs have fully unimpeded access into this region. Consistency with modern data selects the cosmic ray spectra of Hayakawa et al. (1961), Nath and Biermann (1994), and Valle et al. (2002) as most reasonable. For our purposes, the spectrum of Nath and Biermann (1994) produces maximal cosmogenic nuclides, and Valle et al. (2002) produces somewhat less.

Models consistent with Voyager measurements greatly exceed terrestrial cosmic ray fluxes. For cosmic rays with energies greater than the rigidity of the magnetosphere (~10 GeV), the interplanetary and interstellar cosmic ray spectrum remains little changed from the terrestrial level. However, in energy ranges below this, the cosmic ray flux is increased by many orders of magnitude. Of special interest to our work are cosmic rays in the range from 10 MeV – 1 GeV, which are increased 3 to 5 orders of magnitude outside the heliosphere and are also above the threshold for producing cosmogenic nuclides. In particular, cosmic rays of this energy produce abundant secondary neutrons in particle showers, as well as $^{10}$Be and $^{26}$Al through spallation (Lange et al. 1995, Michel et al. 1997).

We model long-period comets as spheres with the composition specified above which spend their lifetime outside the heliopause, before a perturbation sends them into the inner Solar System.

## 4. COSMOGENIC NUCLIDE PRODUCTION

The isotope produced in largest quantities in our atmosphere is $^{14}$C, created by the absorption of cosmic ray produced neutrons by $^{14}$N. Production of $^{10}$Be is caused by the direct spallation of nitrogen and oxygen by protons and neutrons. $^{26}$Al is produced primarily through the spallation of silicon, but other heavy target nuclei produce this isotope as well.

We have performed Monte Carlo simulations using MCNPX 2.6 (Hendricks et al. 2007) which is optimized for neutron transport. Monte Carlo simulators such as MCNPX are extensive programs based on all fundamental interactions with elementary particles, and they are often used for simulation of cosmic ray showers and other high energy interactions. MCNPX has been used to simulate cosmic ray showers in both the atmosphere and in spallation neutron producers. For our simulation, a comet was modeled as a column of ice and dust best fitting what is known of comet nucleus density and composition. It was given a length sufficient for inclusion of all cosmic ray secondaries and periodic boundary conditions. Column composition consisted of equal parts ice and dust, matching the elemental abundances of Halley. Elemental abundances within the ice were chosen to be ~11% hydrogen, ~79% oxygen, ~4% nitrogen, and ~6% carbon by mass (Delsemme 1982). While the dust abundances were chosen to be ~5% hydrogen, ~24% carbon, ~1.5% nitrogen, ~35% oxygen, ~0.6% sodium, ~6% magnesium, ~0.5% aluminum, ~13% silicon, ~6% sulfur, ~0.02% potassium, ~0.6% calcium, ~0.05% titanium, ~0.1% chromium, ~0.07% manganese, ~7% iron, ~0.04% cobalt, ~0.6% nickel (Jessberger et al. 1988). As the density and composition of long-period comets is only partially known, a larger fraction of volatiles was tested in simulations with very little impact on secondary production.

Models were run at 10 logarithmic primary kinetic energies between 10 MeV and 1 GeV, and in order of magnitude bins above 1 GeV. Output was given in the form of cell average flux tallies, including proton and neutron cell average fluxes in 100 logarithmic energy bins. These tallies produce secondary proton and neutron fluxes per primary, which were then convolved with the cosmic ray spectrum to produce total proton and neutron fluxes. As expected, neutron production rates inside the comet were found to exceed atmospheric rates. This is due to the increased density of comet matter, giving secondaries additional interaction opportunities before their decay after ~15 minutes.

Convolution of the acceptable cosmic ray spectra with the individual neutron production rates produced a total neutron flux. This neutron flux was multiplied by the absorption cross section for the nitrogen content within the nucleus to find the rate of $^{14}$C production in the comet. $^{14}$C is radioactive with a half-life of 5730 years (Godwin 1962). The amount of $^{14}$C on the comet at any given time is therefore that sustainable at steady state between production and decay. As we are considering long-period comets, this will provide a very good approximation for the $^{14}$C residing on the comet at any given time. This is because a long-period will spend at most 200 years, a negligible portion of its lifetime, in its first entrance into the heliosphere, which is insufficient for a significant decay of these cosmogenic nuclides. This amount was found by equating the production rate and the decay rate on the comet, as follows:

$$m_C \lambda_C = \int \Phi(m, E, r) \sigma_C(m, E) \, dr \, dE, \qquad (1)$$

where $m_C$ is the mass of $^{14}C$ on the comet, $\lambda_C$ is the decay constant of $^{14}C$ (0.00012 yr$^{-1}$), $\Phi$ is the flux of neutrons in the comet, and $\sigma_C$ is the total interaction cross section for $^{14}C$ production. Both the flux of particles and the interaction cross section will depend on the mass of the comet and the energy of the particles interacting with the comet. The flux of secondary particles depends strongly on the depth of the comet (r) and is reduced to negligible beyond a depth of ~20 m. This is consistent with atmospheric measurements, as the column density of cometary matter at 20 m is equivalent to the column density of the atmosphere at sea level. Integration was done numerically for 20 depth bins of 1 meter each, and 10 order of magnitude neutron energy bins.

The comet will reach this steady state amount during its residence in the interstellar medium (short-period comet rates can be estimated as two orders of magnitude lower). The steady state amount will depend on the cosmic ray spectrum chosen. We have chosen the cosmic ray spectra of Nath and Biermann (1994) and Valle et al. (2002), which produce the range of cosmogenic nuclide production displayed in our results.

$^{10}Be$ is produced through the spallation of oxygen and nitrogen. The interaction cross section for this process has been well studied for a variety of primary energy ranges (Lange et al. 1994, Michel et al. 2007, Kovaltsov and Usoskin 2010). To find the rate of $^{10}Be$ production, we multiplied the proton and neutron fluxes by their corresponding total interaction cross sections. $^{10}Be$ is radioactive with a half-life of 1.387 Myr. To find the amount of $^{10}Be$ present on the comet at any given time, we set the production rate in interstellar medium equal to the decay rate on the comet. This creates a steady state equation similar to that for $^{14}C$:

$$m_{Be} \lambda_{Be} = \sum_i \left( \int \Phi_i(m, E, r) \sigma_{i,Be}(m, E) \, dr \, dE \right), \qquad (2)$$

where $m_{Be}$ is the mass of $^{10}Be$ on the comet, $\lambda_{Be}$ is the decay constant of $^{10}Be$ (~5x10$^{-7}$ yr$^{-1}$), $\Phi_i$ is the flux of high energy particles in the comet, and $\sigma_{i,Be}$ is the total interaction cross section for $^{10}Be$ production for particle type $i$. The flux of high energy particles behaves similarly to the flux of neutrons in equation (1). Numerical integration was used with bins similar to equation (1).

$^{26}Al$ is produced through the spallation of many heavy isotopes, primarily silicon. The interaction cross section for this process has been well studied for a variety of primary energy ranges (Lange et al. 1994, Michel et al. 2007, Kovaltsov and Usoskin 2010). To find the rate of $^{26}Al$ production, we multiplied the proton and neutron fluxes by their corresponding total interaction cross sections. $^{26}Al$ is radioactive with a half-life of 717 kyr. To find the amount of $^{26}Al$ present on the comet at any given time, we set the production rate in interstellar medium equal to the decay rate on the comet. This creates a steady state equation similar to the $^{14}C$ and $^{10}Be$ versions:

$$m_{Al} \lambda_{Al} = \sum_i \left( \int \Phi_i(m, E, r) \sigma_{i,Al}(m, E) \, dr \, dE \right), \qquad (3)$$

where variables are named in analogy to equations (1) and (2), and $\lambda_{Al}$~9.7x10$^{-7}$ yr$^{-1}$.

# 5. RESULTS

## 5.1 Cosmogenic Carbon-14

Comet $^{14}$C abundance has been calculated for comets ranging in size from $10^7$ to $10^{16}$ kg. Figure 1A shows $^{14}$C mass as a function of total comet mass. For this figure, we assume a spherical comet of density between 0.3 g/cm$^3$ and 0.6 g/cm$^3$, and the elemental abundances of Halley as well as solar abundances. The lower bound corresponds to a density of 0.6 g/cm$^3$ and the nitrogen abundance of Halley; the upper bound corresponds to a density of 0.3 g/cm$^3$ and solar nitrogen abundance. This range includes all comet compositions which have been measured.

Any comet small enough to allow secondaries to exit without interaction is also too low in mass to have a measurable impact on the cosmogenic nuclide abundance of the atmosphere. This is due to the secondary neutron flux decreasing to nil after ~20 meters. This makes the mass of cosmogenic nuclide in the relevant mass range go roughly as the surface area, or $m_c^{2/3}$ for a spherical object.

As shown in Figure 1A, the amount of mass deposited in the case of small comets is insignificant compared with the average $^{14}$C mass in the atmosphere, or approximately 500 kg. To make a noticeable change to the $^{14}$C record, an impactor must contain ~1% of this total mass, or ~5 kg. This makes the detection of impactors in $^{14}$C only plausible in the case of very large objects which originally reside primarily outside of the heliosphere. Impact events from such objects are decidedly rare but pose a significant threat to life. Short-period comets and other extraterrestrial objects which orbit primarily within the heliosphere would have cosmogenic nuclide masses almost two orders of magnitude smaller, so they will not be likely candidates for significant $^{14}$C or $^{10}$Be enhancement.

## 5.2 Cosmogenic Beryllium

Comet $^{10}$Be abundance has been calculated for comets ranging in size from $10^7$ to $10^{16}$ kg. Figure 1B displays $^{10}$Be mass as a function of comet mass. We again assume a spherical comet of density between 0.3 g/cm$^3$ and 0.6 g/cm$^3$. A nonspherical cometary nucleus would produce somewhat higher amounts of cosmogenic isotopes.

These results can be compared to experimental results from carbonaceous chondrites. Carbonaceous chondrites are stony meteorites which were tested (Goel 1969) for the presence of $^{10}$Be. The $^{10}$Be concentration was found to be roughly twice what our results show for long-period comets. This difference arises from the lower abundance of (primarily oxygen) target nuclei to create $^{10}$Be on comets, as well as a smaller fraction of comet mass lying within the spallation zone. Although these chondrites are capable of producing cosmogenic nuclides such as $^{10}$Be, they lack the nitrogen content to produce measurable $^{14}$C atmospheric enhancements.

The $^{10}$Be content shown in Figure 1B is large in comparison to present day ice core concentrations. However, as in the case of $^{14}$C, this amount depends greatly on the trajectory and mass of the extraterrestrial object. Unlike $^{14}$C, $^{10}$Be does not depend as heavily on target nitrogen, and thus, provides little information on composition of the extraterrestrial object. Also unlike $^{14}$C, the processes by which $^{10}$Be is deposited into ice cores are dependent on a larger variety of factors, such as snowfall rates and circulation of $^{10}$Be within the atmosphere (Finkel and Nishiizumi 1997). This makes $^{10}$Be ice core abundance less dependent on atmospheric abundance, where cometary $^{10}$Be would be deposited.

**5.3 Cosmogenic Aluminum**

Comet $^{26}$Al abundance has been calculated for comets ranging in size from $10^7$ to $10^{16}$ kg. Figure 1C displays $^{26}$Al mass as a function of comet mass. These results are consistent with measurements of other extraterrestrial matter (Nishiizumi et al. 1995).

The production of $^{26}$Al is greatly enhanced compared to terrestrial levels. This enhancement is caused by substantially larger amounts of silicon being present on the surface of the comet. On the Earth, most of the silicon lies under the protection of a fairly thick atmosphere, so very little $^{26}$Al is made by cosmic rays. Our results show the mass of $^{26}$Al to be near to that of $^{10}$Be. As the deposition processes of $^{26}$Al are very similar to those of $^{10}$Be, the ratio of $^{26}$Al/$^{10}$Be in the ice core would reflect the effect of the ratio on the bolide at the time of deposition. This ratio escapes the uncertainty in deposition processes, making it more dependent on presence of cometary matter in our atmosphere than either one of these isotopes alone.

**6. DISCUSSION**

Our results show that measurable amounts of $^{14}$C will be deposited in our atmosphere by large long-period comets. Short-period comets can be expected to contain two orders of magnitude less $^{14}$C, making them impossible to measure except in bolides larger than ~$5 \times 10^{15}$ kg. Long period comets equal to the catalog average mass should contain a measurable amount of $^{14}$C, up to ~5% of the total atmospheric $^{14}$C or greater, as in the case of a large comet such as Halley. This amount scales as $m_c^{2/3}$, due to the surface area of the comet.

The Tunguska event remains a benchmark for recent bolide impacts, and our work is based on the current estimated $5 \times 10^7$ kg mass of the Tunguska object (Wasson 2003). If the Tunguska impactor were a long-period comet with half its mass in volatiles, the amount of $^{14}$C produced would still be minimal and well below measurable amounts, consistent with the data.

The hypothetical Younger Dryas object is proposed to have been an object with a mass between $4 \times 10^{12}$ kg (Bunch et al. 2012) and $5 \times 10^{13}$ kg (estimated by Toon et al. 1997 as sufficient for continent-wide devastation). Given this range, the object could have deposited $^{14}$C between ~0.5% and ~6% of total atmospheric $^{14}$C. This is sufficient to explain the ~5% $^{14}$C increase measured at the onset of the Younger Dryas event, which has been observed in tree rings (Hua et al. 2009), coral (Stuiver et al. 1998), lake sediments (Ramsey et al. 2012), and ocean sediments (Hughen et al. 2006), as shown in Figure 2. This increase would require a very large mass comet with low density. Impacts of this size are rare. At a rate of $10^{-6}$ per year, admittedly dominated by small-number statistics (Chapman & Morrison 1994) the probability of one in the last 13,000 years is close to 1%. However (Asher et al. 2005; Napier & Asher 2009; Napier 2010) suggest that the rate is much higher, particularly for long-period comet impacts. Although changes in ocean circulation are proposed to affect the $^{14}$C concentration in this record, research suggests that this is not the case (Muscheler et al. 2000). Our results provide an alternative possibility for this deposition, which is currently not well understood.

Another offset in carbon dating is seen at 774 AD (Miyake et al. 2012), and this sudden increase of $^{14}$C is equal to 1.2% of the total $^{14}$C on Earth, which would require a comet roughly 100,000 times more massive than Tunguska. A comet of this size would have caused significant damage near the airburst location and is unlikely because it could have escaped detection only if it occurred far from inhabited areas. Alternately, it has been suggested (Melott and Thomas 2012, Thomas et al. 2013, Usoskin et al. 2013) that this $^{14}$C increase may have been caused by a solar major proton event.

Although our results show that $^{14}$C can be deposited by extraterrestrial impact, an increase in $^{14}$C alone is not conclusive evidence of a past impact. This is due to the other processes which are capable of increasing $^{14}$C, such as geomagnetic reversals and variations in cosmic ray flux. This result does, however, suggest that the $^{14}$C peak at the Younger Dryas onset does not necessitate an additional process aside from bolide impact. This is consistent with previous measurements of Clovis-age sites. These sites contain excess radiocarbon thought to be the result of Younger Dryas impact particles (Firestone 2009), also interpreted as modern contamination (Boslough et al. 2012). We may be seeing either contamination or a contribution of impactor material itself; both could in principle display a negative $^{14}$C age.

Our results also show that measurable amounts of $^{10}$Be will be deposited in our atmosphere by long-period comets. Although the amount of $^{10}$Be could be quite large, measurement of this deposition could be difficult. $^{10}$Be is most often recorded through ice core sampling, and the process controlling transport and deposition of atmospheric $^{10}$Be to ice sheets is not fully understood (Pedro et al. 2011). $^{10}$Be precipitates out of our atmosphere at varying rates depending on geographic location, climate, and other factors. Additionally, $^{10}$Be resides in our atmosphere for less than one year on average (Finkel and Nishiizumi 1997), making detection of an instantaneous event very difficult. For these reasons, we find that a lack of $^{10}$Be signature typically does not rule out a long-period comet impact. For the proposed Younger Dryas comet, the $^{14}$C increase is coeval with a $^{10}$Be peak in ice core data, as shown in Figure 2 (Stuiver et al. 1998, Finkel and Nishiizumi 1997). It has been suggested that this peak may be the result of excess dust accumulation during this time period. However, the increase in $^{10}$Be exists both within the ice core concentration measurements as well as in estimated flux calculations (Finkel and Nishiizumi 1997). Dust accumulation should be increased during the entire Younger Dryas period, which would not explain the sharp increase at the period's onset. An additional effect of the residence time of $^{10}$Be being less than the time required for cross equatorial atmospheric circulation (Finkel and Nishiizumi 1997, Melott and Thomas 2009) is that if the conjectured YD impact were predominantly a northern hemisphere event, then a peak in northern hemisphere ice cores coincident without a peak in southern hemisphere ice cores is expected. Our results show that a long-period comet of the size and impact area hypothesized for the Younger Dryas event would contain enough $^{10}$Be to create this peak, and would deposit $^{10}$Be primarily in the northern hemisphere. Solar proton events or geomagnetic weakening would not show a strong hemispheric asymmetry. Presence or lack of $^{10}$Be alone within ice cores does not provide sufficient evidence for an impact event, due to other factors controlling its deposition process. For this reason, we only use $^{10}$Be data as a consistency check.

$^{26}$Al should also be present in comets or asteroids. The deposition process for $^{26}$Al is believed to be very similar to that of $^{10}$Be. Recent work has focused on the measurement of the $^{26}$Al/$^{10}$Be ratio in ice cores (Auer et al. 2009), and this work found the ratio to remain largely constant through time, but with unexplained increases in older samples of a factor of ~2. The mean ratio of atmospheric $^{26}$Al/$^{10}$Be is

$1.89 \times 10^{-3}$ (Auer et al. 2009). Our results predict the cometary ratio of $^{26}Al/^{10}Be$ to be ~0.6, and other extraterrestrial objects have been measured to have a ratio of 1 or greater. Cosmic spherules from moraines and deep sea sediments have been measured and show increases in the $^{26}Al/^{10}Be$ ratio, up to a ratio of 23 (Nishiizumi et al. 1995). This enhancement over the atmospheric ratio occurs due to the location of silicon target nuclei on or near the exterior of extraterrestrial objects. Most silicon on Earth is protected by the atmosphere, causing $^{26}Al$ production to be much less than in extraterrestrial objects. All comets and asteroids will have an increased $^{26}Al/^{10}Be$ ratio for this reason, regardless of orbital period. Therefore, any comet or asteroid impacting the Earth should increase this ratio, as seen in older ice core samples. Because geomagnetic reversals and increases in cosmic ray flux affect all nuclide production uniformly, neither of these phenomena would create this signature. This ratio has been measured in selected sections of the GISP2 ice core[2], but no measurements have been made around the time of the Younger Dryas event. Under current conditions, ~5% of the total $^{26}Al$ deposited in ice cores comes from extraterrestrial mass (Auer et al. 2009). An impactor of the size hypothesized for the Younger Dryas event would be ~$10^4$-$10^6$ times more massive than the yearly average extraterrestrial mass flux, and therefore, would inject many times the average yearly $^{26}Al$ mass, and deposition of this large amount of $^{26}Al$ would change the ratio of $^{26}Al/^{10}Be$ at that time. Assuming the peak in $^{10}Be$ from GISP ice cores at the Younger Dryas event is from deposition of extraterrestrial matter, the ratio of $^{26}Al/^{10}Be$ should increase by a factor of ~100+ at that time. The $^{26}Al/^{10}Be$ ratio is ~1 for all extraterrestrial matter (Auer et al. 2009), and therefore would increase even if the Younger Dryas impactor was not a long-period comet. As $^{10}Be$ is strongly deposition dependent, and 14C increases can be associated with a variety of phenomena, this ratio proves to be a much better test of the presence of extraterrestrial cosmogenic nuclides within our atmosphere. This ratio should be measured with sufficient resolution for $^{26}Al$ detection as a test for the Younger Dryas impact hypothesis and would be applicable regardless of impactor type.

    We have found that the amount of $^{10}Be$ and $^{14}C$ produced in airbursts or impacting long-period comets should be detectable in the geologic record, which can be used to rule out a class of large events. Of known and conjectured cometary airbursts, the Younger Dryas comet remains the only impactor of requisite size for measurable $^{10}Be$ and $^{14}C$ deposition. We find that the increases in $^{14}C$ and $^{10}Be$ at the time of the Younger Dryas are consistent with the airburst of a long-period comet of the suggested mass, making this a plausible scenario. This is not a conclusive test of whether the event occurred, but rather demonstrates the consistency of the hypothetical Younger Dryas object producing this effect. To properly test the Younger Dryas impact hypothesis, $^{26}Al/^{10}Be$ ratios must be measured around the time of the purported event. A short lived increase in this ratio would be compelling evidence of a large impact at this time by an extraterrestrial object. Lack of a short lived increase of this ratio from a complete sampling would be inconsistent with a large extraterrestrial impact. We note that due to the

---

[2] Nishiizumi, K., Finkel, R. C., and Welten, K. C., 2005, 26Al in GISP2 Ice Core, *in* Proceedings, The

    10th International Conference on Accelerator Mass Spectrometry, September 5-10, 2005,

    Berkeley, California.

expected large increase in $^{26}$Al abundance for a large meteor, the required mass of the ice sample is much lower than normally used to examine usual terrestrial ratios. We estimate that the mass of ice required for this measurement would be 5 to10 kg, as opposed to ~100 kg.

## 7. CONCLUSIONS

We propose a method by which abundance peaks in cosmogenic nuclides in the terrestrial geologic record can be used to detect past ET impacts. We show that measurable increases in terrestrial records of $^{14}$C and $^{10}$Be are expected from large long-period comets but not significant from short-period comets or asteroids. Carbon sphereules and other residue directly associated with the impact may show even larger relative increases in $^{14}$C fraction. Regarding the proposed Younger Dryas impact event, we calculate that the magnitudes of known $^{14}$C and $^{10}$Be peaks are consistent with the impact of a long-period cometary body within the limits of the previously proposed mass, but are not conclusive of such an impact since they could have other causes. On the other hand, large-scale deposition of $^{26}$Al is characteristic of any large extra-terrestrial impactor. Therefore, we recommend future high-resolution analyses of $^{26}$Al/$^{10}$Be ratios in ice cores and ocean cores across the Younger Dryas boundary at 12.9 ka as a way to further test the Younger Dryas impact hypothesis. Such analyses could also detect other previously unknown impact events.

## ACKNOWLEDGMENTS

We are grateful for discussions with Michael Murray in the early stages of this project. Research support at the University of Kansas was provided by NASA Program Astrobiology: Exobiology and Evolutionary Biology under grant number NNX09AM85G.

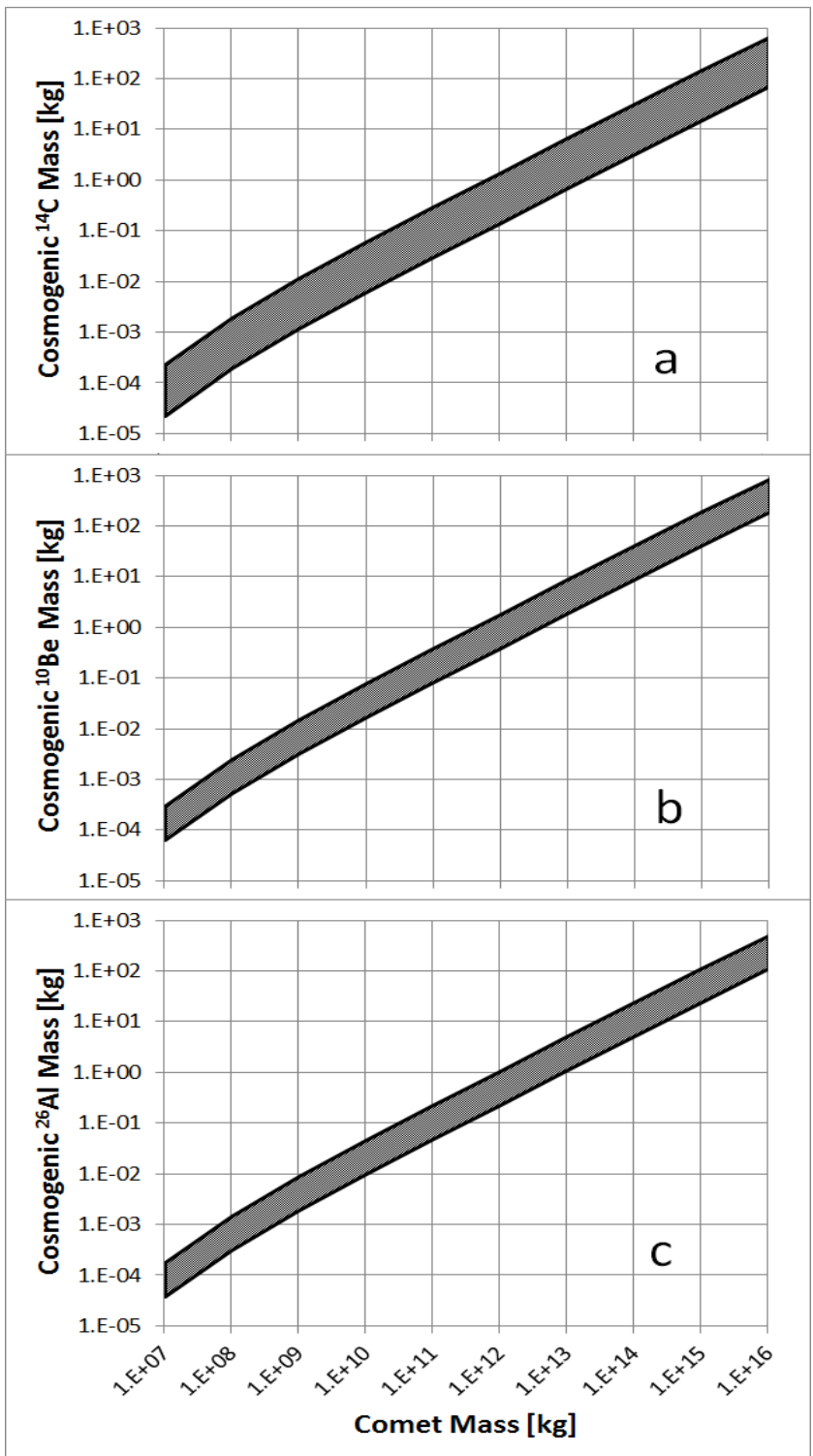

Figure 1. Cosmogenic $^{10}$Be, $^{14}$C, and $^{26}$Al mass contained on a long-period comet as a function of comet mass. Shaded regions designate uncertainty based on the interstellar medium cosmic ray spectrum, comet density, and composition. Figure 1A corresponds to cosmogenic carbon, Figure 1B corresponds to cosmogenic beryllium and Figure 1C corresponds to cosmogenic aluminum. All axes are plotted logarithmically.

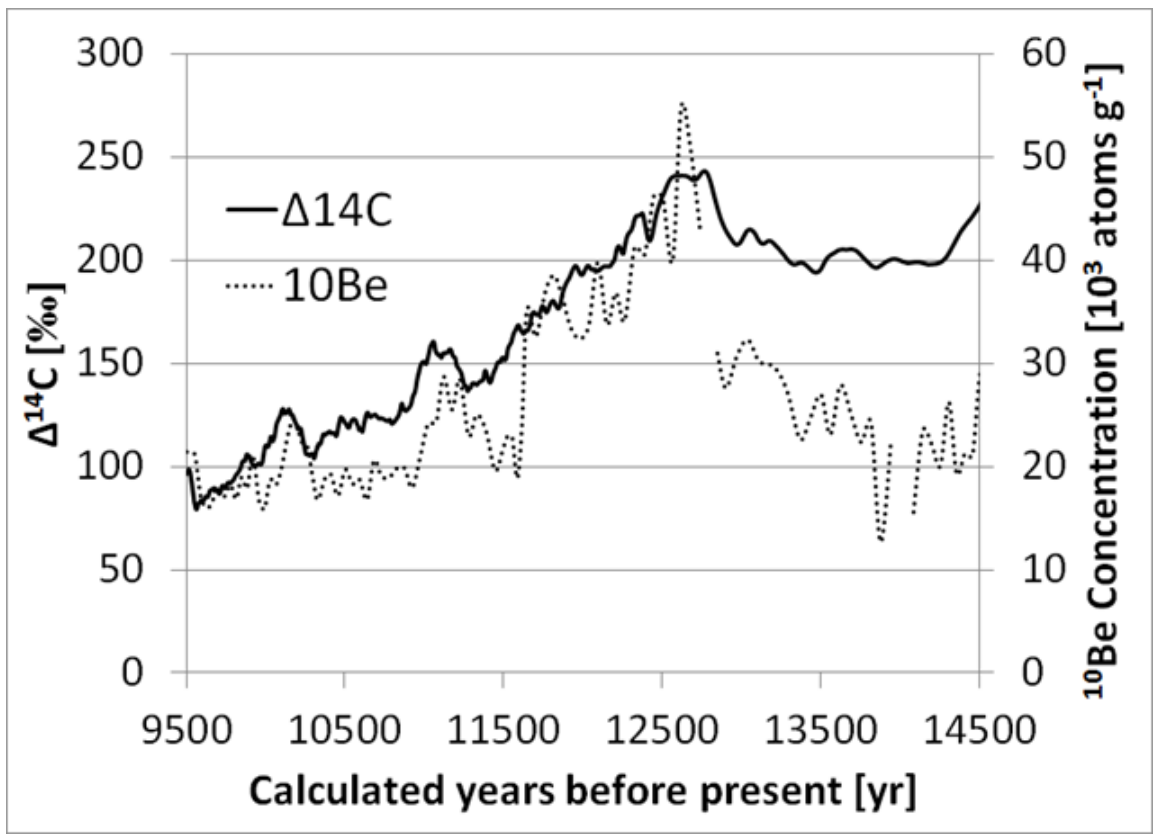

Figure 2. $\Delta^{14}C$ and $^{10}Be$ concentration measured from 9500 to 14500 years before present. The $\Delta^{14}C$ is measured per mil from tree rings, coral, and marine sediment and is shown with a solid line, from Stuiver et al. 1998. The $^{10}Be$ concentration in thousands of atoms per gram of ice is measured from GISP ice cores and is shown with a dotted line, from Finkel and Nishiizumi 1997. This figure displays the sudden increase in both $\Delta^{14}C$ (~5%) and $^{10}Be$ concentration (~80%) close to the beginning of the Younger Dryas event, at 12900 years before present.